\documentclass[twocolumn,amsmath,amsfonts,amssymb]{revtex4}
\usepackage{graphicx}
\def\beq{\begin{equation}}
\def\eeq{\end{equation}}
\def\bea{\begin{eqnarray}}
\def\eea{\end{eqnarray}}

\begin{document}

\title{Entropic force, running gravitational coupling and future singularities}

\author{Maryam Aghaei Abchouyeh}
\email{ m.aghaei@ph.iut.ac.ir}
\affiliation{Department of Physics, Isfahan University of Technology, Isfahan, 84156-83111, Iran}
\author{Behrouz Mirza}
\email{b.mirza@cc.iut.ac.ir}
\affiliation{Department of Physics, Isfahan University of Technology, Isfahan, 84156-83111, Iran}
\author{Zeinab Sherkatghanad}
\email{zsherkat@uwaterloo.ca}
\affiliation{Department of Physics, Isfahan University of Technology, Isfahan, 84156-83111, Iran\\
Department of Physics and Astronomy, University of Waterloo, Waterloo, Ontario, Canada, N2L 3G1}

\begin{abstract}
The effects of a running gravitational coupling and the entropic force on future singularities are considered. Although it is expected that the quantum corrections remove the future singularities or change the singularity type, treating the running gravitational coupling as a function of energy density  is found to cause no change in the type of singularity but causes a delay in the time that a singularity occurs. The entropic force is found to replaces the singularity type $II$ by $\overline{III}$ ($a=$const., $H=$const., $\dot{H} \to \infty$, $p \to \infty$,  $\rho \to \infty$  ) which differs from previously known type $III$ and to remove the $w$-singularity. We also consider an effective cosmological model and show that the types $I$ and $II$ are replaced by the singularity type $III$.
\end{abstract}
\maketitle


\section{Introduction}
Both the theoretical cosmology and observational data
indicate that our accelerated expanding universe can be described by an equation of state (EOS) parameter
$w$ around $-1$ \cite{w,N1,S1}. In general, when the universe passes
through the $\Lambda$CDM epoch, $w$ is exactly $-1$. However, a phantom-dominated universe
can be described by $w$ slightly less than $-1$ while if $w$ is
slightly more than $-1$, the quintessence dark epoch occurs.
In the region where barotropic index $w$ is lower than $-1$, a future singularity may occur in the
form of an infinite scale factor, energy density, and pressure which has come to be called the
"big rip".  Other types of singularities have been explored, increasing the family
of candidates \cite{clasify3,N}. It is known that
 different types of singularities may arise during the expansion of the universe,
which may be classified as follows \cite{clasify1,clasify2,ww,dabro,denki,K}\\
Type $I$ (Big Rip): Infinite $a$, $\rho$, and $p$.\\
Type $II$ (Sudden): Finite $a$, $H$, and $\rho$; divergent $\dot{H}$ and $p$.\\
Type $III$ (Big Freeze): Finite $a$; infinite $H$, $\rho$, and $p$. This singularity type is a subcase of Finite Scale Factor singularity\\
Type $IV$: Finite $a$, $H$, $\dot{H}$ , $\rho$, and $p$ but infinite higher derivatives of $H$.\\
Type  $V$: Infinite $w$ (barotropic index): By expanding the scale factor around the time
of the singularity, we get  $a(t)=c+(t_{s} -t)^{n_1}+(t_{s} -t)^{n_2}+....$, where $n_1 <n_2 <....$ and $c>0$ \cite{w2,clasify4,w1}. In this way,  $w$-singularity may occure without any divergence in pressure or energy density with the following properties: \\
(i)\ $p_{s} \neq 0$: $n_1 = 2$ and $n_2 = [3, \infty)$,\\
(ii)\ $p_{s}= 0$: $n_1 = [3, \infty)$ and $n_2 = [n_1 +1, \infty)$.\\
$w$-singularity was first proposed in \cite{w1} with a different form of the scale factor.\\
\indent According to  Tipler's definition \cite{Ti}, the big rip (Type $I$) singularity  is strong whereas types $II$, $III$, $IV$, and $V$
 are weak. Based on Krolak's definition \cite{Kr}, however, only  types $II$, $IV$, and  $V$  are weak \cite{clasify5}.\\
Conformal anomaly effects can change the type of singularity but bulk viscosity merely decreases the time when singularities may happen \cite{anomaly,S,viscous1,viscous2}.\\
\indent  We may adopt an approach in which the concepts of information and holography play the central role \cite{eforce1,eforce1a,eforce2,eforce3,barrow,Cai:2010zw,Cai:2010kp}. By holography is meant a situation in which  all the information of a volume $V$ can be encoded on its boundary screen. In this situation, the effect of entropic force can produce the generalized Friedman equations to yield a new type of future singularity. The entropic force will change the singularity from Type $II$ to $\overline{III}$  ($a=$const., $H=$const., $\dot{H} \to \infty$, $p \to \infty$,  $\rho \to \infty$) which differs from previously known type $III$.\\
In general, the gravitational coupling is assumed as a universal constant, but the idea of variation of physical constants such as the tational coupling
$G$, the charge of the electron $e$, the velocity of
light $c$ has been investigated in both theoretical and experimental physics \cite{Ca,BM,Cb,Cd,Ce,RG1,RG2,RG3,cc2,G1,G2,MH} and might be effective to avoid singularities \cite{cc1}. In this paper we study an asymptotically safe scenario as a quantum effect to avoid some exotic behaviors of the universe near the singularity time. Considering  Gravitational coupling as a function of energy density in the asymptotically safe scenario \cite{BM,as,Wein} does not change the type of future singularity but only delay  the time while a singularity appears.\\
\indent This paper is organized
as follows: Sec. II investigates the singularities of the generalized Friedman
equations with a running gravitational coupling. Sec. III examines the effect of entropic force on future singularities. The effects of both running gravitational coupling and entropic force on the different types of singularities are studied in Section IV. In Sec. V,  an effective  cosmological model is used to study future singularities. It is expected that a quantum correction removes the predicted future singularities of the universe or at least changes the singularity type.

\section{Running gravitational coupling and the future singularities}
\label{sec:1}
\begin{figure}[h]
\centering
\includegraphics[width=0.85\textwidth]{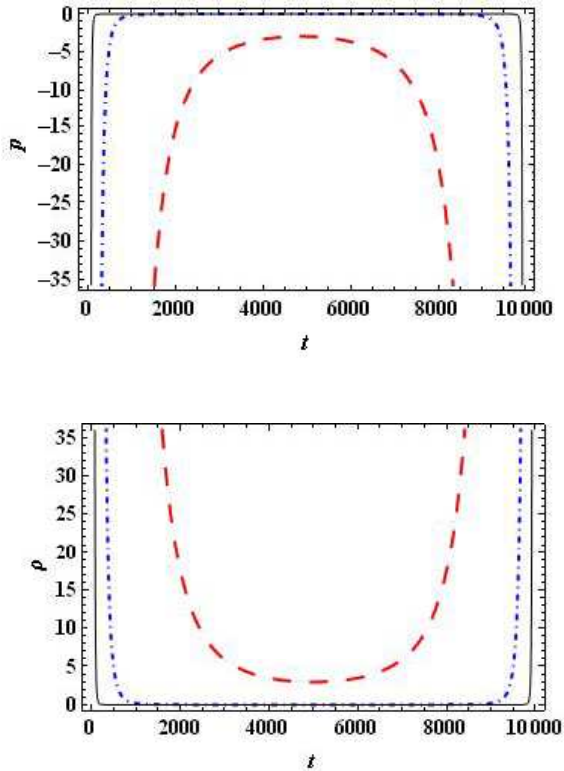}
\caption[]{\it{Pressure $p$ and energy density $\rho$ in an asymptotically safe scenario with respect to
   $t$ for  Type I singularity for $n=3$, $t_{s}=10000$, and $\beta=0.5$ [red (dashed) line], $0.6$ [blue (dashed-dotted) line], and
   $0.7$ [black (solid) line].}} \label{fig:figure 1}
\end{figure}

 Following \cite{Ca,cc1}, we consider a Running gravitational coupling varying by time. The generalized Friedman equations are given by
\begin{eqnarray}
 \label{1}
    &&3H^{2}=8\pi G(t) \rho,\\
    &&\frac{\ddot{a}(t)}{a(t)}=\frac{-4\pi G(t)}{3}(\rho +3\frac{p}{c^2}),
\end{eqnarray}
here $a(t)$ is the scale factor and the dots is the divertive with respect to time. In this condition the Bianchi identity holds with respect to the covariant derivative of each side of the Einstein field equation in this way
\begin{eqnarray}
 \label{1a}
  \dot{\rho}+3H(\rho +\frac{p}{c^2})=-\rho \frac{\dot{G}(t)}{G(t)}.
\end{eqnarray}
A running gravitational coupling may represent some essential features of an asymptotically safe gravitational model \cite{BM}. An asymptotically safe gravitational model is important because of its quantum background. It is expected that some quantum corrections on General Relativity can remove the predicted future singularities of the universe or at least change the singularity type. In an asymptotically safe model the gravitational coupling $G$ should be a  function of an energy scale, so the more usual choice energy scale is energy density and the energy density is a function of time. Thus the gravitational coupling can be taken as $G(\rho) \equiv G(\rho (t))$,
where $G(\rho(t))\sim\rho(t)^{-\frac{\alpha}{2+\alpha}}$ and $\alpha\geq2$, which represent a kind of asymptotically safe model
\cite{BM}.

Therfore, the pressure $p(t)$ and energy density
$\rho(t)$ for this form of running gravitational coupling
are given by ($c=1$)
\begin{eqnarray}
       \label{2}
    &&\rho(t)=[\frac{3}{8\pi} H^{2}(t)]^{\frac{1}{1-\beta}},\\
     \label{2a}
     &&p(t)=-\frac{1}{4\pi}(\frac{3}{2} H^{2}(t)+\frac{d H(t)}{dt})[\frac{3}{8\pi}H^{2}(t)]^{\frac{\beta}{1-\beta}},
     \end{eqnarray}
where, $\beta=\frac{\alpha}{2+\alpha}$ and, for
asymptotic safety $\beta\geq0.5$.\\
\indent Now let us consider the effect of running gravitational coupling on future singularities. Since the energy density is divergent in  Type $I$ and $III$ singularities, we can consider the effect of a running gravitational coupling on these types of future singularities.
In general, the scale factor $a(t)$ associated with
Type $I$  for the standard Friedman equations  can be expressed as follows \cite{clasify1}
 \begin{equation} \label{3}
    a(t)=a_{0}(\frac{t}{t_{s}-t})^{n},
 \end{equation}

\begin{figure} 
 \includegraphics[width=0.5\textwidth]{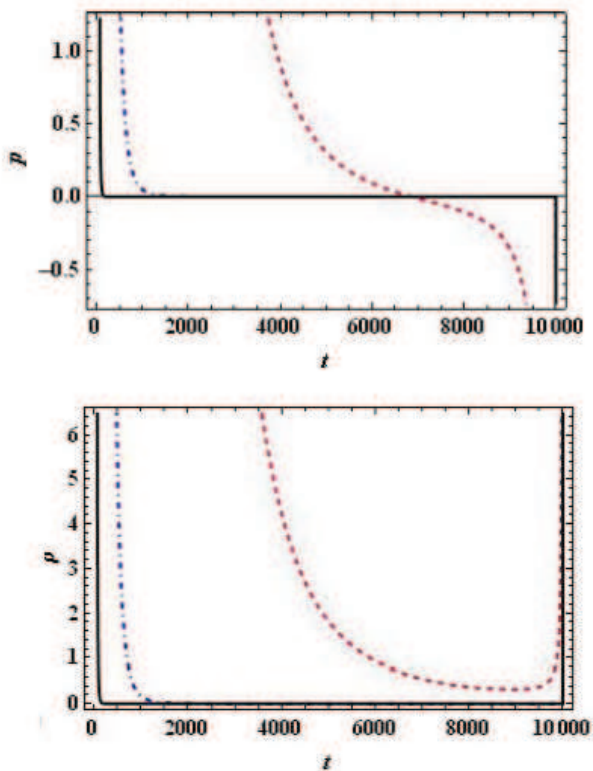}
\caption{$p$ and $\rho$  with respect to
   $t$ (asymptotically safe scenario) Type III singularity for $n=0.5$, $q=0.5$, $a_s=10$, $t_{s}=10000$ and $\beta=0.5$ [red (dashed) line], $0.6$ [blue (dashed-dotted) line], and
   $0.7$ [black (solid) line].}
\label{fig:figure 2}
\end{figure}

\noindent where, $n$ is a positive constant and $0 < t < t_{s}$. In this situation, the scale
factor diverges at a finite time $(t\rightarrow t_{s})$ where
$t_s$ is the instant of singularity. For the scale factor in Eq. (\ref{3}), the behaviors of the
energy density and pressure in Eqs. (\ref{2}) and (\ref{2a}), respectively, show that Type $I$  does not change in response to the effect of a running gravitational coupling (Fig. \ref{fig:figure 1}). This type of singularity is so strong that it cannot be affected by an asymptotically safe  scenario; however, the larger values of $\beta$ near the singularity causes a delay in  the time  the singularity appears.\\

\begin{figure} 
  \includegraphics[width=0.87\textwidth]{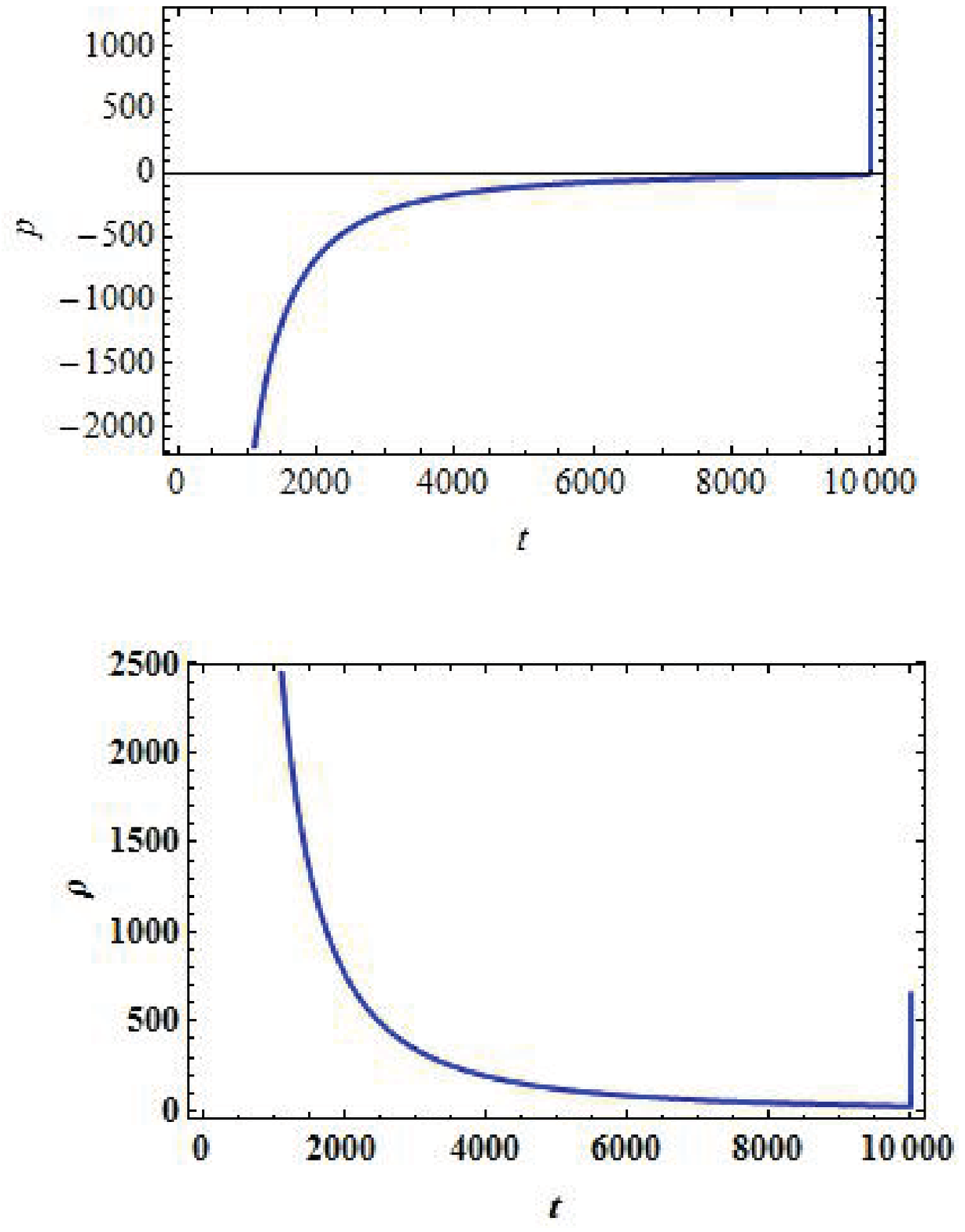}
\caption{Pressure $p$ and energy density $\rho$ with respect to
   $t$ (entropic force scenario) for Type $\overline{III}$ singularity for $t_{s}=10000$, $q = 0.5$, $n = 1.5$, $c_1=0.1$, $c_2=0.01$.}
\label{fig:figure 3}
\end{figure}

\indent The scale factor related to Type
$III$ singularity is given as follows \cite{denki,barrow}:
\begin{equation}\label{4}
    a(t)=1-(1-\frac{t}{t_s})^{n}+(\frac{t}{t_s})^{q}(a_{s}-1),
\end{equation}
For the special values $n<1$ and $0<q<1$, Type $III$ singularity occurs and the strong and weak
energy conditions are violated.  Therefore, Type $III$ singularity should remain intact by considering a running gravitational coupling in an asymptotically safe scenario. However,  our results show that larger values of $\beta$ or a weaker gravitational force for the singularity  types $I$ and $III$   correspond to a larger value of $t_s$, which means a delay in the appearance of   the singularities. Thus, we may expect that an asymptotically safe  running gravitational coupling (as a quantum effect) does not change the singularity types nor remove any of the singularities. Pressure $p(t)$ and energy density
$\rho(t)$ are depicted in  Fig. \ref{fig:figure 2}.

\section{The effect of entropic force on the future singularities }
\label{sec:2}

In this Section, we consider the effect of  entropic force on  future singularities.
Easson, Frampton and Smoot obtained the modified Friedman equations by considering the entropic force scenario, which introduces entropic force as a result of surface effects \cite{eforce1a,eforce2,eforce3}. Thus, the Friedman equations are generalized to the following form \cite{eforce1a}
  \begin{eqnarray} \label{5}
       &&\frac{\ddot{a}(t)}{a(t)}=-\frac{4\pi G}{3}(\rho+3p)+c_1H^{2}(t)+c_2\dot{H}(t)\\
       &&H^{2}(t)=\frac{8\pi G}{3}\rho +c_1H^{2}(t)+c_2\dot{H}(t),
     \end{eqnarray}
Here, the coefficients  $c_1$ and $c_2$ are determined by
observations \cite{eforce2,eforce3}. In this way,  energy density $\rho(t)$ and
pressure $p(t)$ are given by
\begin{eqnarray}
    \label{6}&&\rho(t)=\frac{3}{8\pi G}[H(t)^{2}-c_1H(t)^{2}-c_2\dot{H}(t)],\\
    \label{6a}&&p(t)=\frac{1}{4\pi G}[\frac{-3+3c_1}{2}H(t)^{2}+\frac{-2+3c_2}{2}\dot{H}(t)].
     \end{eqnarray}

\begin{figure} 
\center
\includegraphics[width=0.5\textwidth]{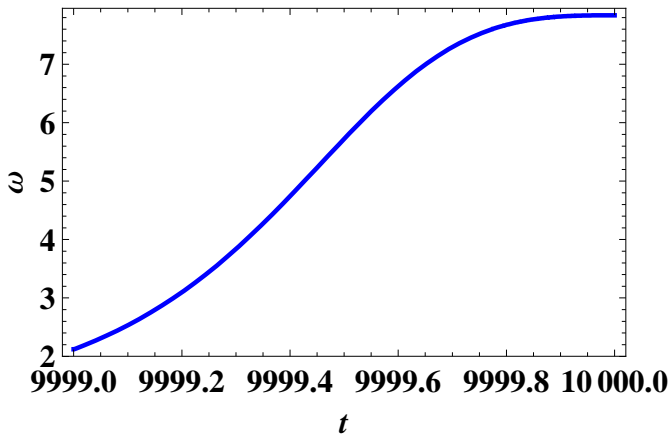}
\caption{$w$ with respect to
   $t$ (entropic force scenario) for $n=2$, $q=3.5$, $c = 4$, $c_1=0.2$, $c_2=0.01$, and $t_{s}=10000$.}
\label{fig:figure 4}
\end{figure}

\begin{table*}
\caption{Violated energy conditions for
$q=0.5$, $n=1.5$ in an entropic force scenario.}
\label{sphericcase}
\begin{tabular*}{\textwidth}{@{\extracolsep{\fill}}lrrrrrl@{}}
\hline
$c_1$ & $c_2$ & $\frac{c_1}{c_2}$ & Violated energy conditions & Singularity\\
\hline
$0<c_1<1$  & $0<c_2<1$  & $>1$ & DEC and DNEC & II $\rightarrow$ $\overline{III}$\\
$0<c_1<1$  & $0<c_2<1$  & $<1$ & No & II $\rightarrow$ $\overline{III}$\\
$c_1>1$  & $c_2>1$ & $>1$ & No & II $\rightarrow$ $\overline{III}$\\
$c_1>1$  & $c_2>1$ & $<1$ & No & II $\rightarrow$ $\overline{III}$\\
$c_1>1$  & $0<c_2<1$ & $>1$ & No & II $\rightarrow$ $\overline{III}$\\
\hline
\end{tabular*}
\end{table*}
Now let us consider the effect of entropic force on the future singularities. For the scale factor related to Type $I$  in Eq. (\ref{3}), we can plot $\rho(t)$ and $p(t)$ in Eqs. (\ref{6}) and (\ref{6a}) with respect to $t$. The results indicate no change for Type $I$ singularity. Also, the sudden singularity (Type $II$) occurs for the scale factor represented in Eq. (\ref{4}) when $1<n<2$ and $0<q<1$
and the dominant energy condition is violated \cite{anomaly}.
Using the scale factor for Type $II$ singularity and the energy density $\rho(t)$ and
pressure $p(t)$ in Eqs. (\ref{6}) and (\ref{6a}) respectively, we find that the Hubble parameter $H$ is finite although $\dot{H}$, $\rho$, and $p$ are infinite. $\rho$ diverges as a result of $\dot{H}$ term in Eq. (\ref{6}). Thus, for all positive values of $c_1$ and $c_2$ a new type of singularity is taking place which we call it type $\overline{III}$ (Fig. \ref{fig:figure 3}). It is similar to Type $III$  but only with a different behavior of $H$. This new type can be characterized by  finite $a$ and $H$ as well as infinite $\dot{H}$, $\rho$, and $p$. Turned into  Krolak's definition, a weak singularity (Type $II$) turns into a strong singularity (Type $\overline{III}$) \cite{Ti}. Furthermore, the violated energy conditions are sensitive to the values of coefficients $c_1$ and $c_2$. The results are presented in Tables  I and II.\\
\indent For Type $IV$ and $w$-singularity, the scale factor is given by
\cite{ww}
\begin{equation}
\label{4a}
    a(t)=c+(t_s-t)^n+(t_s-t)^q,
\end{equation}
\noindent where, $c$ is a positive constant. The special values of $2\leq n <\infty$ and
$n<q<\infty$ correspond to the singularity Type $IV$. Also the $w$-singularity occurs when: i) $n= 2$ and $q= [3, \infty)$, ii) $n= [3, \infty)$ and $q= [n+1, \infty)$.\\
\indent Our results also show that the singularity Types $III$ and $IV$ which are described by the scale factors in Eq. (\ref{4}) and Eq. (\ref{4a}), respectively, remain intact. It is interesting that the effect of entropic force removes the $w$-singularity  (Fig. \ref{fig:figure 4}). It should be noted that the presence of $\dot{H}$ in Eqs. (\ref{6}) and (\ref{6a}) is responsible for the removal of the $w-$singularity in entropic force scenario.

\begin{figure} 
 \includegraphics[width=0.5\textwidth]{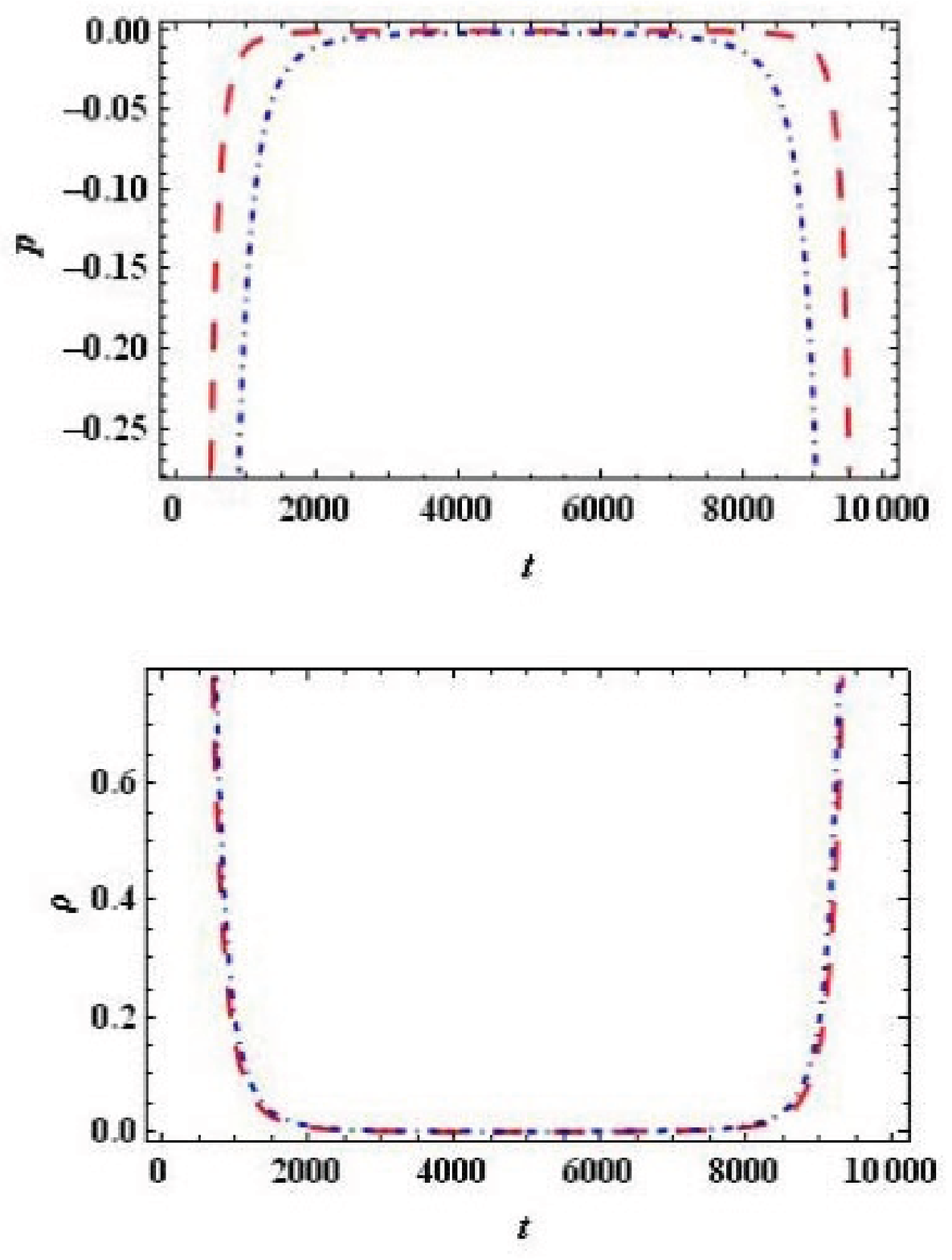}
\caption{$p$ and $\rho$ with respect to
   $t$ for the singularity Type $I$
   $t_s = 10000$, $n=3$ by considering the effect of a running gravitational coupling $\beta=0.6$, $c_1=c_2=0$ [blue (dashed-dotted) line] and by the effects of both the entropic force and running gravitational coupling $\beta=0.6$, $c_1=0.1$, $c_2=0.01$ [red (dashed) line].}
\label{fig:figure 5}
\end{figure}

\section{ The effects of both the entropic force and running gravitational coupling on future singularities}
\label{sec:3}

In the previous sections, we showed that the running gravitational coupling does not change the type of singularity. However, the effect of entropic force changes the singularity type from  $II$ to  $\overline{III}$. Now, we account for the effects of both running gravitational coupling and entropic force on Friedman equations
\begin{eqnarray}
    \label{7b}
    &&\frac{\ddot{a}(t)}{a(t)}=-\frac{4\pi}{3}\rho ^{-\beta}(\rho+3p)+c_1H^{2}(t)+c_2\dot{H}(t)\\
     \label{7c}
     &&H^{2}(t)=\frac{8\pi}{3}\rho ^{1-\beta} +c_1H^{2}(t)+c_2\dot{H}(t).
     \end{eqnarray}
Thus, Eqs. (\ref{6}), (\ref{6a}) reduce to
\begin{eqnarray}
    \label{7}
    &&\rho(t)=[\frac{3}{8\pi }(H(t)^{2}-c_1H(t)^{2}-c_2\dot{H}(t))]^{\frac{1}{1-\beta}},\\
     \label{7a}
     &&p(t)=\frac{1}{4\pi}(\frac{-3+3c_1}{2}H(t)^{2}+\frac{-2+3c_2}{2}\dot{H}(t)) \\\nonumber
     &&\times [\frac{3}{8\pi }(H(t)^{2}-c_1H(t)^{2}-c_2\dot{H}(t))]^\frac{\beta}{1-\beta}.
    \end{eqnarray}
Our asymptotically safe scenario is not relevant to the singularity types $II$, $IV$, and $V$  as the energy density remains finite at the time the singularity appears. Using the scale factor in Eq. (\ref{3}) for the singularity type $I$ and Eq. (\ref{4}) for  type $III$, we can plot the energy density $\rho(t)$ and pressure $p(t)$ in Eqs. (\ref{7}) and (\ref{7a}) with respect to time. In this case, neither Type $I$ nor Type $III$ change as a result of  the effects of either
running gravitational coupling or entropic force. These types  are classified as  strong singularities along the lines of Krolak's definition \cite{Ti,Kr}.\\
Our results indicate that larger values of $\beta$ that represent a weaker gravitational force for the singularity types $I$ and $III$  still correspond to a larger value of $t_s$ or a delay in the time that the singularities appear and further that the entropic force increases this effect of running gravitational coupling (Fig. \ref{fig:figure 5}). This is a novel behavior and characterizes the effects of both running gravitational coupling and entropic force on singularity types $I$ and $III$.

\section{An effective cosmological model }
\label{sec:4}

In this section, we consider an effective cosmological model which is a map from the generalized equations to the standard Friedman equations. Thus,
Eqs. (\ref{7b}) and (\ref{7c}) can be expressed as
 \begin{equation} \label{8}
     H_{eff}^2(t)=\frac{8\pi}{3} \rho ,
 \end{equation}
in compare with Friedmann Equations in general relativity, where,\\
 \begin{equation}\label{8a}
 H^2 _{eff}(t)=\frac{8\pi}{3}[\frac{3}{8\pi }(H(t)^{2}-c_1H(t)^{2}-c_2\dot{H}(t))]^\frac{1}{1-\beta}.
 \end{equation}
  For the effective scale factor,  we have
  \begin{equation} \label{9}
     a_{eff}(t)=\exp[\int H_{eff}(t)] dt,
  \end{equation}
Aditionally, using Eq. (\ref{8}) and the conservation law $\dot{\rho}_{eff}+3 H_{eff}(t)
(\rho_{eff}+p_{eff})=0$, we have
 \begin{equation} \label{10}
     \frac{\ddot{a}_{eff}(t)}{a_{eff}(t)}=-\frac{4\pi}{3} (\rho_{eff}+3p_{eff}), \\
  \end{equation}
where
  \begin{eqnarray} \label{11}
     &&\rho _{eff}=\frac{3 H_{eff}^2(t)}{8\pi} \\
     \label{11a}
     &&p_{eff}=-\frac{1}{4 \pi}(\frac{3}{2} H_{eff}^2(t)+ \dot{H}_{eff}(t)) ,
  \end{eqnarray}
Here, effective energy density and pressure have their original definitions $\rho(t)\equiv \rho_{eff}(t)$ and $p(t)\equiv p_{eff}(t)$, respectively in Eqs. (\ref{11})and (\ref{11a}). This means that by a redefinition for Hubble parameter the original form of Friedmann Equations remain unchanged. In order to study which singularities may change under this new definitions of cosmological parameters we consider Eqs. (\ref{9}), (\ref{11}) and (\ref{11a}). Using, the scale factor associated with Type  $I$ singularity in Eq. (\ref{3}), we have shown that, for $c_1, c_2 \geq 0$, the effective scale factor in Eq. (\ref{9}) is finite although $H_{eff}$, $\rho$ and $p$ are infinite. Thus, Type $I$ singularity is replaced by  Type ${III}$. The singularity type ${III}$  is defined as finite $a_{eff}$ and infinite $H_{eff}$, $\rho$ and $p$. In other words, considering an asymptotically safe gravity and the entropic force lead to the replacement of the strong  singularity of Type $I$ with the weaker one of type ${III}$ in the effective cosmological model.\\
\indent Let us consider the effective scale factor in Eq. (\ref{9}) for the other future singularities. For the singularity type $II$  the scale factor $a(t)$ described by Eq. (\ref{4}) and the effective scale factor by Eq. (\ref{9}). our results indicate that $H_{eff}$, $\rho$ and $p$ are infinite for $\beta=0$ and $c_1, c_2 > 0$. So, Type $II$ singularity is replaced by Type ${III}$. Furthermore, the singularity type $III$  in Eq. (\ref{4}) for $\beta \geq 0.5$ as well as types $IV$ and $V$ in Eq. (\ref{4a}) for $\beta=0$ remain the same as before. The results are summarized in Table II.

\begin{table}
\centering
\caption{The effects of  entropic force and running gravitational coupling on  future singularities.}
\label{parset}
\begin{tabular*}{\columnwidth}{@{\extracolsep{\fill}}llll@{}}

\hline
Entropic Force  & $c_1 =0.1$, $c_2 =0.01$ & $II$ $\rightarrow$ $\overline{III}$ \\
                \hline
An effective  & $c_1 =0.1$, $c_2 =0.01$, $\beta \geq 0.5$ & $I$ $\rightarrow$ ${III}$ \\
cosmological & &\\
\ model & $c_1 =0.1$, $c_2 =0.01$, $\beta=0$ & $II$ $\rightarrow$ ${III}$\\
\hline
\end{tabular*}
\end{table}

\section{Conclusion}
\label{sec:5}
In this paper, we studied the effects of running gravitational coupling and  entropic force on future singularities.
Since the energy density is divergent for types $I$ and $III$ in the singularity time, we can consider the effect of asymptotically safe running gravitational coupling for these types of future singularities. The effect of running gravitational coupling does not change the type of singularity. However, for singularity types $I$ and $III$, a weaker gravitational force (larger values of $\beta$) corresponds to a delay in the time that the singularity appear.\\
\indent Furthermore, the entropic force  changes the  singularity from Type $II$  to a new type  called Type $\overline{III}$ ($a=$const., $H=$const., $\dot{H} \to \infty$, $p \to \infty$,  $\rho \to \infty$) which differs from previously known type $III$. The  type of singularity is similar to type $III$ but only with a different behavior of $H$. Although the singularity types $I$, $III$ and $IV$  do not change as a result of  considering the effect of entropic force, Type $V$, however, is removed in this case.\\
\indent  Our results indicated that both the entropic force and a running gravitational coupling cause a delay in the singularity types $I$ and $III$. Finally, by introducing a dual cosmological model, we investigated different types of singularity. It was shown that types $I$ and $II$ change to type ${III}$. It is interesting that Type $I$ is replaced by a weaker one along the lines defined by Tipler. It should be noted that the singularity types $III$, $IV$ and $V$  do not change in this dual cosmological model.



\end{document}